\documentclass{iau} 
\usepackage{natbib,graphicx}

\renewcommand{\[}{\begin{equation}}
\renewcommand{\]}{\end{equation}}

\def\ex#1{\left\langle#1\right\rangle}
\def\msun{{\rm\,M_\odot}}
\def\kpc{\,{\rm kpc}}\def\kms{\,{\rm km\,s^{-1}}}
\def\spose#1{\hbox to 0pt{#1\hss}}
\def\lta{\mathrel{\spose{\lower 3pt\hbox{$\mathchar"218$}}
     \raise 2.0pt\hbox{$\mathchar"13C$}}}
\def\gta{\mathrel{\spose{\lower 3pt\hbox{$\mathchar"218$}}
     \raise 2.0pt\hbox{$\mathchar"13E$}}}

\let\boldgrk=\gkvecten
\let\boldgrksc=\gkvecseven

\def\gkthing#1{{\mathchoice%
	{\hbox{{\boldgrk\char#1}}}
	{\hbox{{\boldgrk\char#1}}}
	{\hbox{{\boldgrksc\char#1}}}
	{\hbox{{\boldgrksc\char#1}}}}}

\def\vtheta{\gkthing{18}}

{\newif\ifnotend
\notendtrue
\def\veclist{ABCDEFGHIJKLMNOPQRSTUVWXYZabcdefghijklmnopqrstuvwxyz.}
\def\top#1#2.{#1}
\def\tail#1#2.{#2.}
\loop\expandafter\xdef\csname v\expandafter\top\veclist\endcsname%
{{\noexpand\bf\expandafter\top\veclist}}
\edef\veclist{\expandafter\tail\veclist}
\if\veclist.\notendfalse\fi\ifnotend\repeat}
\def\d{{\rm d}}

\title[Symposium 353.~~Galactic Dynamics in the Era of Large Surveys] 
{Modelling our Galaxy}

\author[James Binney]   
{James Binney}

\affiliation{Rudolf Peierls Centre for Theoretical Physics, University of
Oxford, Parks Road, Oxford OX1 3PU, UK \\ email: {\tt binney@physics.ox.ac.uk}}

\pubyear{2019}
\volume{353}  
\jname{Galactic Dynamics in the Era of Large Surveys}
\editors{M. Valluri, \& J. A. Sellwood  }
\begin{document}

\maketitle

\begin{abstract}
Dynamical models will be key to exploitation of the incoming flood data for
our Galaxy. Modelling techniques are reviewed with an emphasis on $f(\vJ)$
modelling.
\end{abstract}

\firstsection 
\section{Introduction}

Since the second data release by the Gaia project in 2018 April we have been
in possession of an enormous body of precision data for our very typical
Galaxy. From satellites and ground-based telescopes we have astrometry and
photometry (Gaia, Kepler, WISE, 2-Mass, SkyMapper, PanSTARRS) and spectra,
largely taken from the ground, at sufficient resolution to infer chemical
compositions (APOGEE, RAVE, LAMOST, Gaia-ESO). Data of increased precision
for more stars will continue to arrive over the next $\sim$5 years. The task
that lies before us is to use these data to understand how our Galaxy is
structured, how it functions as a machine, and how it was assembled. Those
tasks need to be addressed in roughly that order: it is pointless to
speculate how the Galaxy was assembled until it is clear how it is
structured, and not much more profitable to speculate about assembly until
you understand how its dynamics causes its structure to evolve over cosmic
time.

Dynamical models will be central to understanding how the Galaxy is
structured. In the 1980s a significant step forward in our understanding of
the Galaxy was taken when inversion of star counts \citep[e.g.][]{Bok1937} was
replaced by forward modelling with star counts predicted by a sum of
components -- exponential disc, $r^{1/4}$ bulge, power-law stellar halo --
derived from observations of external galaxies
\citep{BahcallS1980,CaldwellO1981}. For
many years, the de facto standard Galaxy model, the Besan\c con model
\citep{Roea03}, has been a development of this idea, in which the kinematics of
stars are pasted onto a model derived from star counts. That is, at each
location the distribution of stellar velocities is assumed to be bi- or
tri-axial Gaussian with pre-determined principal axes. In reality the
velocity distributions of a given component at different locations are
coupled by dynamics. 

The traditional approach to acknowledging this coupling
is the use of Jeans' moment equations, which relate the moments $\nu\ex{v^n}$
(with $\nu$ a component's stellar number density) to the Galactic
gravitational field $-\nabla\Phi$. There are many objections to this
procedure. A basic one is that the Jeans' equation for $\nu\ex{v^n}$ involves
$\nu\ex{v^{n+1}}$ and an ansatz is required to close this hierarchy. Another
problem is that velocity distributions are always significantly non-Gaussian
and important information is contained in the deviations from Gaussianity.
Consequently, the higher moments $\nu\ex{v^n}$ for $n>2$ are independent of
the moments for $n=1$ (mean velocity) and $n=2$ (velocity dispersion), and
adequate characterisation of the system 
requires values for the higher moments. It's hard to recover them reliably
from  the Jeans
equations \citep{MagorrianB1994}.

\section{Modelling methods}

\subsection{$f(\vJ)$ modelling}

A promising approach to model building starts not with each component's density
$\nu(\vx)$ but with its distribution function (DF) $f(\vx,\vv)$. Once the DFs
of every component is known, the system is completely specified in the sense
that one can easily predict the outcome of any measurement; the set of
DFs for a galaxy's components is the galaxy's DNA. 

Models with specified DFs exploit Jeans' theorem, which states that the DF of
a steady-state system can be assumed to depend on $(\vx,\vv)$ only through
the constants of motion $I_k(\vx,\vv)$.
Specification of a stellar system by writing down a DF goes back to the
origins of stellar dynamics \citep{Eddington1916,Jeans1916} and it has played a major
role in studies of globular clusters \citep{Henon1960,King1966}. Until
recently its application to galaxies was largely to spherical galaxies
\citep{Jaffe1983,He90} and razor-thin discs \citep{Kalnajs1977}, despite
brave efforts to apply it to S0 galaxies
\citep{PrendergastT1970,Rowley1988}.
All these early efforts were restricted to systems with single components, a
serious restriction given the importance of dark matter and gradients in
chemical composition and stellar age within galaxies: serious engagement with
observational data requires multi-component models, which predict how colour
varies through a system and relates this variation to velocity dispersions
measured in different spectral lines.

In a steady-state system, energy $E$ is always a constant of motion, and all
classical models make the DF a function of $E$. The key to multi-component
modelling is to resist the temptation to put $E$ into the DF's argument list.
The reason is that the energy of an orbit is not a local quantity. For
example, the simplest orbit is that on which the star sits at the galactic
centre. If we add a shell of dark matter, the energy of this orbit will be
depressed, but the orbit remains the same. When we write down the DF of the bulge or the stellar halo, we will want to specify the
phase-space density of that component at this simplest orbit. But until the
model is fully assembled and its potential has been determined, we don't know
the energy $E_0=\Phi(0)$ of this orbit. So we can't specify $f(E,..)$ to
begin model construction.

Much the best constants of motion to use as arguments of $f$ are the action
integrals $J_r$, $J_\phi$ and $J_z$. The first of these quantifies the
amplitude of oscillations in radius (eccentricity); in an axisymmetric system
the second is simply the angular momentum around the symmetry axis;  the last
$J_z$
quantifies the amplitude of oscillations perpendicular to the equatorial
plane. In addition to having simple physical interpretations, action
integrals are adiabatic invariants: if the potential evolves over many
dynamical times (perhaps because accretion of gas has built up a massive
stellar disc), orbits evolve such that their actions remain the same.
Consequently, $f(\vJ)$ is invariant under slow growth of a galaxy.

There should be as many independent actions as the system has spatial
dimensions, and each action can be complemented by a canonically conjugate
`angle' variable $\theta_i$ such that the set $(\vtheta,\vJ)$ comprises a
complete set of canonical coordinates for phase space. The equations of
motion of these variables are trivial: clearly half the equations read $\d
J_i/\d t=0$, while the other three equations read $\d\theta_i/\d
t=\Omega_i(\vJ)$, a constant. Hence the angle variables increase linearly
with time at the rate given by the frequencies $\Omega_i$. The way the system
responds to perturbations is largely determined by how the frequencies
$\Omega_i$ vary with actions, so these frequencies are of fundamental
importance.

For decades the use of action integrals was impractical for lack of
algorithms to translate between $(\vx,\vv)$ and $(\vtheta,\vJ)$. Over the
past decade the situation has improved hugely although it is still not
entirely satisfactory. Excellent approximations to $\theta_i(\vx,\vv)$ and
$J_i(\vx,\vv)$ for orbits in
axisymmetric systems can be obtained from the St\"ackel Fudge
\citep{JJB12:Stackel}, in which formulae strictly valid for St\"ackel
potentials are extended to general potentials. \cite{AGAMA} released
very efficient code (AGAMA) to
implementing  the axisymmetric Fudge.
\cite{SaJJB15:Triaxial} extended the
St\"ackel Fudge to triaxial potentials with negligible figure rotation
rates.

A radically different approach provides the inverse map:
$[\vx(\vtheta,\vJ),\vv(\vtheta,\vJ)]$. This is `torus mapping'. The surface
specified by holding constant $\vJ$ and varying $\vtheta$ is a three-torus in
six-dimensional phase space. Separation of the Hamilton-Jacobi equation for
some potentials of high symmetry yields  analytic expressions
$\vx(\vtheta',\vJ')$ for such tori,
and the Torus Mapper \citep{JJBPJM16} computes the generating function
$S(\vtheta',\vJ)$ that maps these tori into the orbital tori
$\vx(\vtheta,\vJ)$ of any specified axisymmetric potential.

It turns out that simple functional forms $f(\vJ)$ generate models that
closely resemble classical models. \cite{JJBPJM11:dyn} defined a
(`quasi-isothermal') DF that has exponential dependence on the actions that
has been used in several studies to represent exponential discs
\citep[e.g.][]{JJB12:dfs,BoRi13,Piea14}. \cite{Poea15} gave a
form that self-consistently generates double-power-law components including
the NFW and Hernquist spheres. \cite{Pascale2018,Pascale2019} gave a form that
self-consistently generates objects that resemble dwarf spheroidal galaxies
and globular clusters. 

Using these pre-defined DFs a model of a composite system like our Galaxy is
quickly generated: a DF $f_k(\vJ)$ is specified for the dark halo and each stellar
component. The mass $M_k$ of each component is  specified by the
normalisation chosen or $f_k$ because $M_k=(2\pi)^3\int\d^3\vJ\,f_k$. Then
a rough guess $\Phi_0$ is made of the galaxy's total potential and this guess
is used to evaluate the density $\rho_k(\vx)$ of each component on a grid of
locations $\vx$. From these densities and Poisson's equation one recovers a
new estimate of the total potential $\Phi_1$, which is used to re-evaluate
the densities, so an improved potential $\Phi_2$ can be recovered. This
sequence of densities and potentials reliably converges after $\sim5$
iterations. Then the model is complete and ready to predict {\it any}
observable.

\subsection{Schwarzschild modelling}

An orbital torus $[\vx(\vtheta,\vJ),\vv(\vtheta,\vJ)]$ is a powerful extension
of a numerically integrated orbit $[\vx(t),\vv(t)]$. \cite{Schwarzschild1979}
introduced the idea of assembling galaxies by weighting orbits computed in a
trial potential, and this idea has been used extensively to model elliptical
and spheroidal galaxies \citep[e.g.][]{vdBea08,Zhu2018}. Replacing orbits
(time series) with orbital tori has many advantages:

\begin{itemize}
\item One can quickly find the velocity $\vv$ at which a torus will visit a
given location $\vx_0$, while a finite time series  is unlikely to include
$\vx_0$;
\item It's much easier to ensure that a torus library, rather than an orbit
library, samples phase space
systematically and with pre-defined density;
\item A torus is typically specified by less than 100 numbers, whereas an
orbit requires tens of thousands of numbers for its specification;
\item An infinity of new tori can be obtained by interpolating between the
tori of a grid;
\item  Angle-action variables are the language of Hamiltonian perturbation
theory, so a torus-based model can be used to determine how a galaxy responds
to perturbations \citep{MonariDF2016,Binney2018}.
\end{itemize}

I believe that in Schwarzschild modelling tori provide superior replacements
for any quasi-periodic orbit that is a member of the principal families
identified by \cite{Schwarzschild1979} and \cite{deZeeuw1985}. As I'll explain
below, orbital tori can also replace resonantly trapped orbits. The extent to
which they can replace chaotic orbits is still an open question.

\subsection{N-body modelling}

N-body models are enormously flexible and pretty easy to make. Over the last
half century N-body codes have grown enormously in speed and precision, to
the point that a good basic model can be run on a quality laptop and it's
practical to produce a suite simulations with tens of millions of particles
on a small cluster.  

A key issue is the level of Poisson (shot) noise. The
Galaxy's dark halo contains $\sim10^{12}\msun$ and the Poisson noise to
which it gives rise heats a realistically thin disc at an unacceptable rate
if it is represented by fewer than a few million particles
\citep[e.g.][]{Aumer2016a}. Our observational data constrain the disc within
$\lta1\kpc$ from the Sun, a region that contains only $\sim0.24\%$ of the
disc's stars. Hence with $10^6$ particles representing the disc, comparisons
with data will be based on $\sim2400$ phase-space points. This number is too
small given that one needs to examine changes with $|z|$ and split the disc into components by age and
chemistry: stars young and old, metal-poor and metal-rich, low or high in
$\alpha$-abundance, etc.

Another key issue with N-body models is initial
conditions: they specify the equilibrium model that you get after integrating
for a few dynamical times. If the initial conditions lie far from
equilibrium, as in a simulation of cosmological clustering, adjusting the
initial conditions is not a viable way to steer the model towards conformity
with observational data. Hence cosmological simulations don't offer a
promising way to construct a model of our Galaxy as we find it now.  Rather
they are a tool with which to explore, in a general way, how galaxies like
ours might have formed. 

An N-body model is specified by  $\gta10^8$ numbers, the phase-space
coordinates of its particles. With every timestep, all these numbers change while an
equilibrium model does not. Hence the numbers are highly redundant and the
model is needlessly cumbersome. We might be able to gain insight into the
essential structure of an N-body model by fitting its particle distribution
to DFs $f(\vJ)$ for a disc, a bulge, and stellar and dark halos.
Such a fit would simultaneously reduce the model to a few dozen numbers and
give physical insight into its structure. 

Perhaps the most promising way to initialise an N-body model of our Galaxy is to draw
initial conditions from an equilibrium model constructed by the $f(\vJ)$
technique -- the AGAMA package provides a fast routine for furnishing
initial conditions. N-body models initialised in this way provide the best
available tool with which to investigate the collective dynamics of discs:
since the unperturbed model is accurately in equilibrium, time-dependent
features can be reliably ascribed to whatever perturbation one has applied.
Moreover, analytic understanding of why the model responds as it does is
facilitated by the knowledge of the equilibrium DF as a function of actions.
The prospects for gaining an adequate understanding of spiral structure and
warps by this route seem bright.

\subsection{Made-to-measure modelling}

By far the best current model of the Galactic bulge/bar is that constructed by
\cite{Wegg2015} by the made-to-measure (M2M) technique. This may be
considered a hybrid of the N-body and Scharzschild techniques: one starts
with an N-body model and progressively modifies the weights with which orbits
contribute to the overall density and observables so as to optimise the fit
the model provides to observations. Effectively, the N-body model provides
the orbit library, while a `force for change' \citep{SyerTremaine} replaces linear
or quadratic programming as the algorithm for choosing orbit weights. The
technique can model any type of galaxy and is relatively easy to implement.
Its main drawback is that, like any N-body model, it is specified by
$\sim10^8$ numbers that individually lack significance.

\begin{figure}
\includegraphics[width=\hsize]{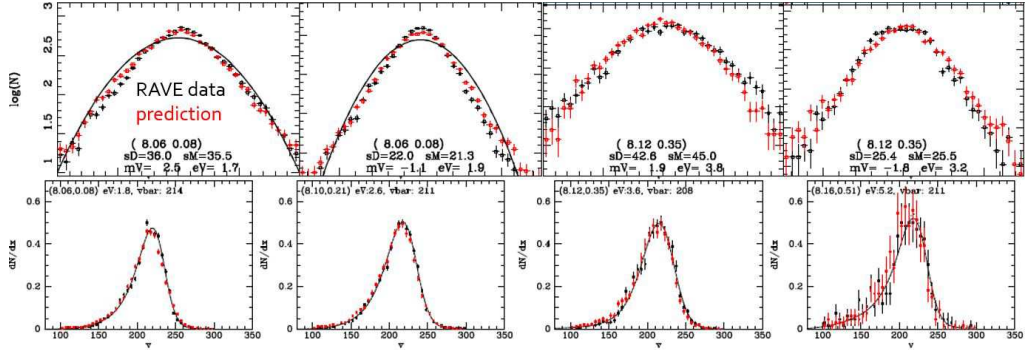}
\caption{Velocity distributions of red dwarfs from the RAVE survey (black points) compared
to predictions (red) of DFs fitted to data from the GCS survey. The top row
shows histograms for $V_R$ and $V_z$ at $\sim R_0$ near the plane 
(left two panels) and at $|z|\sim0.35\kpc$. The lower row shows $V_\phi$ distributions
at distances from the plane that increase from left to
right with the furthest bin centre at  $0.51\kpc$.}\label{fig:RAVE}
\end{figure}

\begin{figure}
\includegraphics[width=\hsize]{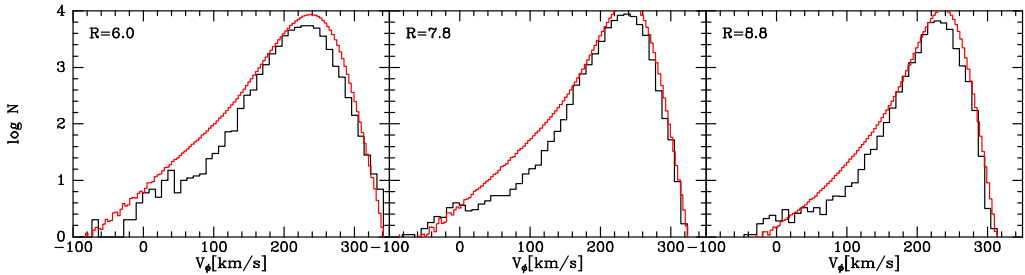}
\caption{Distributions of $V_\phi$ a locations near the plane and radii
$R=6$, $7.6$ and $8.8\kpc$  from  RVS data (black) and
the predictions of a fully self-consistent model for bin centres (red). The peaks of the histograms
are determined by the circular-speed curve of the recovered
potential shown in Fig.~\ref{fig:DR2b}.}\label{fig:DR2a}
\end{figure}

\begin{figure}
\centerline{\includegraphics[width=.6\hsize]{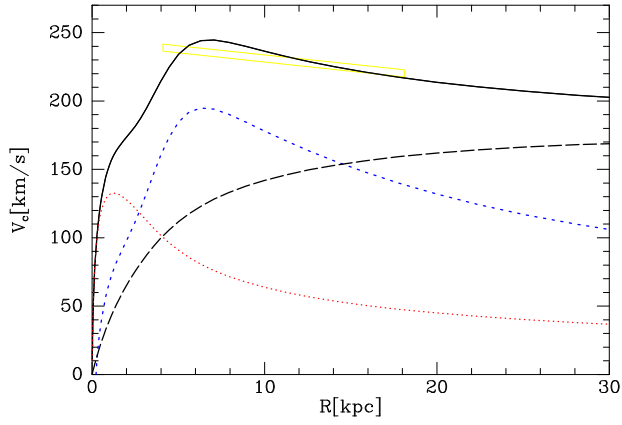}}
\caption{Circular speed curve of a self-consistent Galaxy model fitted to
Gaia DR2 RVS data. The broken curves show the contributions from the dark
halo (long dashed) disc (short dashed) and the bulge
(dotted). The yellow shaded region shows the probable zone of $V_c$ inferred
from Cepheids by \cite{Mroz2019}}\label{fig:DR2b}
\end{figure}

\section{Examples of $f(\vJ)$ modelling}

\subsection{Our Galaxy}

\cite{JJB12:dfs} fitted quasi-isothermal DFs for the thin and thick discs
to data from the Geneva-Copenhagen survey \citep{GCS04,GCS09}.  In figures like
Fig.~\ref{fig:RAVE},
\cite{BinneyBurnett} compared the {\it predictions} of these DFs to the newly
available data from the RAVE survey \citep{RAVE2_short}. The agreement between data (black) and prediction (red)
is spectacular. The parabolas in the two panels at top left show Gaussian
distributions and one sees that the model reproduces the
non-Gaussianity of the data, just as it reproduces the skewness of the
$V_\phi$ distributions shown in the lower panels.

In work prior to 2015 the dark halo was specified by a density distribution
rather than a DF. \cite{PifflPenoyreB} opened a new chapter by specifying the dark
halo through its DF, and \cite{BinneyPiffl15} fitted such a model to data from
several sources, including RAVE and terminal velocities from HI and CO
observations. The heterogeneous nature of their data and limitations of the
software available to them made the fitting process tortuous and costly. Gaia
DR2 data now extend over a sufficiently large radial range that one can dispense
with terminal velocities and determine the structure of the dark halo from
stars alone. Fig.~\ref{fig:DR2a} shows an example: the black histograms are
velocity distributions from the RVS sample in DR2 binned in real space using
the distances of \cite{SchoenrichME2019}, while red curves show a model's
predictions for the centre of the bin. The location of the peak in $V_\phi$
is determined by the circular-speed curve of the derived potential.
Fig.~\ref{fig:DR2b} shows that curve in black together with the contributions
to it from the dark halo (long-dashed), disc (short-dashed) and bulge
(dotted).

\begin{figure}
\centerline{\includegraphics[width=.7\hsize]{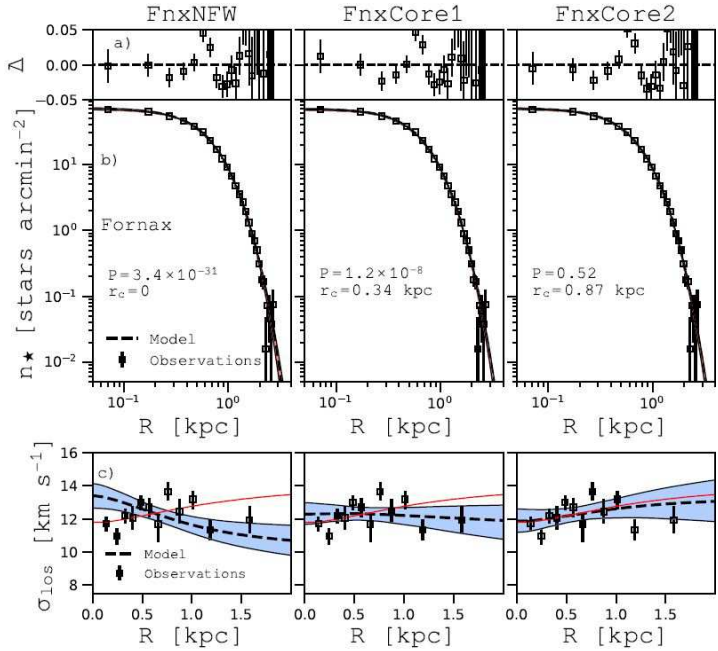}}
\caption{Star counts (upper row) and velocity dispersions (lower row) in the
Fornax dSph compared to predictions of three models. Left column: a model in which the
dark halo has a DF that in isolation generates an NFW profile; middle and
right columns: models in which the dark halo's DF generates a
core.}\label{fig:Pascale}
\end{figure}

\subsection{The Fornax dwarf spheroidal galaxy}

Since stars currently contribute little to the gravitational potentials of
dwarf spheroidals, it has often been argued that these systems have pristine
dark halos in which we can test the prediction of dark-matter-only
cosmological simulations that such halos should have cuspy cores.
\cite{Pascale2018} built $f(\vJ)$ models of the Fornax dSph in which the dark
halo was specified by a DF that either did, or did not, generate a central cusp.
The upper row of Fig.~\ref{fig:Pascale} shows that excellent fits to the
star counts could be obtained with both types of halo DF, but the lower row
shows that the data for velocity dispersion unambiguously favour a DF that
does not generate a cusp.

\begin{figure}
\centerline{\includegraphics[width=.7\hsize]{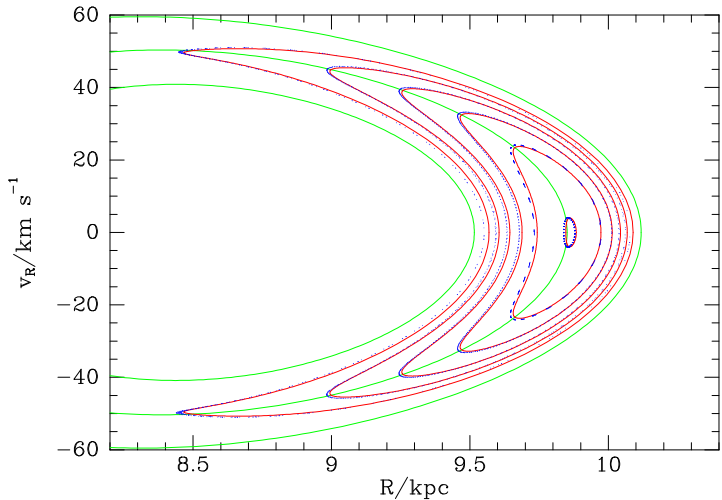}}
\caption{A surface of section for an  axisymmetric Galactic potential in a
region of resonant trapping. Consequents of numerically integrated orbits are
shown in blue, with cross sections of tori computed with the Torus Mapper in
magenta. The green curves show the untrapped tori used by the Torus Mapper.
\citep[From][]{Binney2016}}\label{fig:TM}
\end{figure}

\section{Resonant trapping}

Even in an axisymmetric potential, it's not possible to assign values of
$J_r$, $J_z$ and $J_\phi$ to every orbit. Some orbits become `trapped' by a
resonance. Fortunately, the Torus Mapper provides tools that yield
astonishingly accurate analytic
representations of the tori of trapped orbits such as those shown in
Fig.~\ref{fig:TM}.
\cite{Binney2018} showed that the same tools can be used to obtain analytic
representations of the orbital tori of rotating barred potentials. Trapped
orbits constitute a large topic that I haven't space to explore. From the
perspective of $f(\vJ)$ modelling, they are a complication because each
family of trapped orbits has it's own angle-action coordinates and requires
its own DF $f(\vJ)$. Trapped orbits permit kinematics that can differ
qualitatively from what would be possible in the absence of trapping, such as
circulation in the $(R,z)$ plane \citep{JJBPJM16}. The discovery that at
$|V_z|\sim80\kms$ more (thick disc) stars are passing up/down through the
plane than down/up would provide a valuable constraint on $\Phi$ by  locating
the prominent $\Omega_r=\Omega_z$ resonance.

\bibliographystyle{mn2e} \bibliography{/u/tex/papers/mcmillan/torus/new_refs}

\end{document}